\documentclass[fleqn,usenatbib]{mnras}
\usepackage{amsmath}
\usepackage{graphicx}
\usepackage{amssymb,txfonts}

\newcommand{\suzaku}{{\textit{Suzaku}}}
\newcommand{\nustar}{{\textit{NuSTAR}}}

\title[Anisotropy of partially self-absorbed jets]{Anisotropy of partially self-absorbed jets and the jet of Cyg X-1}

\author[A. A. Zdziarski et al.]
{Andrzej A. Zdziarski,$^1$ Debdutta Paul,$^2$ Ruaraidh Osborne$^3$ and A. R. Rao$^2$\\
$^1$Nicolaus Copernicus Astronomical Center, Polish Academy of Sciences, Bartycka 18, PL-00-716 Warszawa, Poland\\
$^2$Tata Institute of Fundamental Research, Mumbai 400005, India\\
$^3$School of Physics and Astronomy, University of Glasgow, Glasgow G12 8SU, UK\\
}

\date{Accepted 2016 August 17. Received 2016 August 2; in original form 2016 June 10}

\pagerange{\pageref{firstpage}--\pageref{lastpage}}
\pubyear{2016}

\begin{document}

\maketitle

\label{firstpage}

\begin{abstract}
We study the angular dependence of the flux from partially synchrotron self-absorbed conical jets (proposed by Blandford \& K{\"o}nigl). We consider the jet viewed from either a side or close to on axis, and in the latter case, either from the jet top or bottom. We derive analytical formulae for the flux in each of these cases, and find the exact solution for an arbitrary angle numerically. We find that the maximum of the emission occurs when the jet is viewed from top on-axis, which is contrast to a previous result, which found the maximum at some intermediate angle and null emission on-axis. We then calculate the ratio of the jet-to-counterjet emission for this model, which depends on the viewing angle and the index of power-law electrons. 

We apply our results to the black-hole binary Cyg X-1. Given the jet-to-counterjet flux ratio of $\gtrsim$50 found observationally and the current estimates of the inclination, we find the jet velocity to be $\gtrsim 0.8c$. We also point out that when the projection effect is taken into account, the radio observations imply the jet half-opening angle of $\lesssim 1\degr$, a half of the value given before. When combined with the existing estimates of $\Gamma_{\rm j}$, the jet half-opening angle is low, $\ll 1/\Gamma_{\rm j}$, and much lower than values observed in blazars, unless $\Gamma_{\rm j}$ is much higher than currently estimated.
\end{abstract}
\begin{keywords}
acceleration of particles--galaxies: jets--radiation mechanisms: non-thermal--radio continuum: stars--stars: individual: Cyg~X-1--stars: jets.
\end{keywords}

\section{Introduction}
\label{intro}

The radio emission of jets in the hard state of black-hole binaries and of extragalactic parsec-scale radio sources is often flat in the ${\rm d}F/ {\rm d}E$ representation, $\propto E^\alpha$, with $\alpha\sim 0$, where $E$ is the photon energy, e.g., \citet{cawthorne91}, \citet{fender00}, \citet{healey07}. The emission of this type of jet is usually interpreted as due to the partially self-absorbed synchrotron process in a continuous conical jet (\citealt{bk79}, hereafter BK79). The emission at a given energy is self-absorbed from its onset up to a certain height of $z\propto E^{-1}$, and it is optically thin at higher $h$. The jet becomes optically thin at all $z$ at energies above the synchrotron break energy, $E_{\rm t}$.

The model of BK79 is also assumed in an important method of measuring magnetic fields of extragalactic jets from core shifts, i.e., angular displacements with frequency of the observed maxima of jet radio emission \citep{lobanov98,pushkarev12}. This method can also be used to measure the jet power \citep{zamaninasab14, Zdziarski15}.

The angular emissivity pattern of such continuous, partially optically-thick, jets differs from that of optically-thin jets, either steady-state or in the form of moving blobs \citep{lb85,sikora97}. The case of the angular dependence of the emission from partially self-absorbed jets was considered by \citet{cawthorne91}. Here, we re-examine this problem, and obtain different results at viewing angles close to the jet axis. The cause of this discrepancy is that the usual approximation that we view the jet from a side in the comoving frame breaks down in that case.

We also consider implications of our results for the ratio of the jet-to-counterjet emission. We then apply our results to the compact radio jet observed from Cyg X-1 in its hard spectral state.

\section{Definitions}
\label{def}

\begin{figure}
\centerline{\includegraphics[width=\columnwidth]{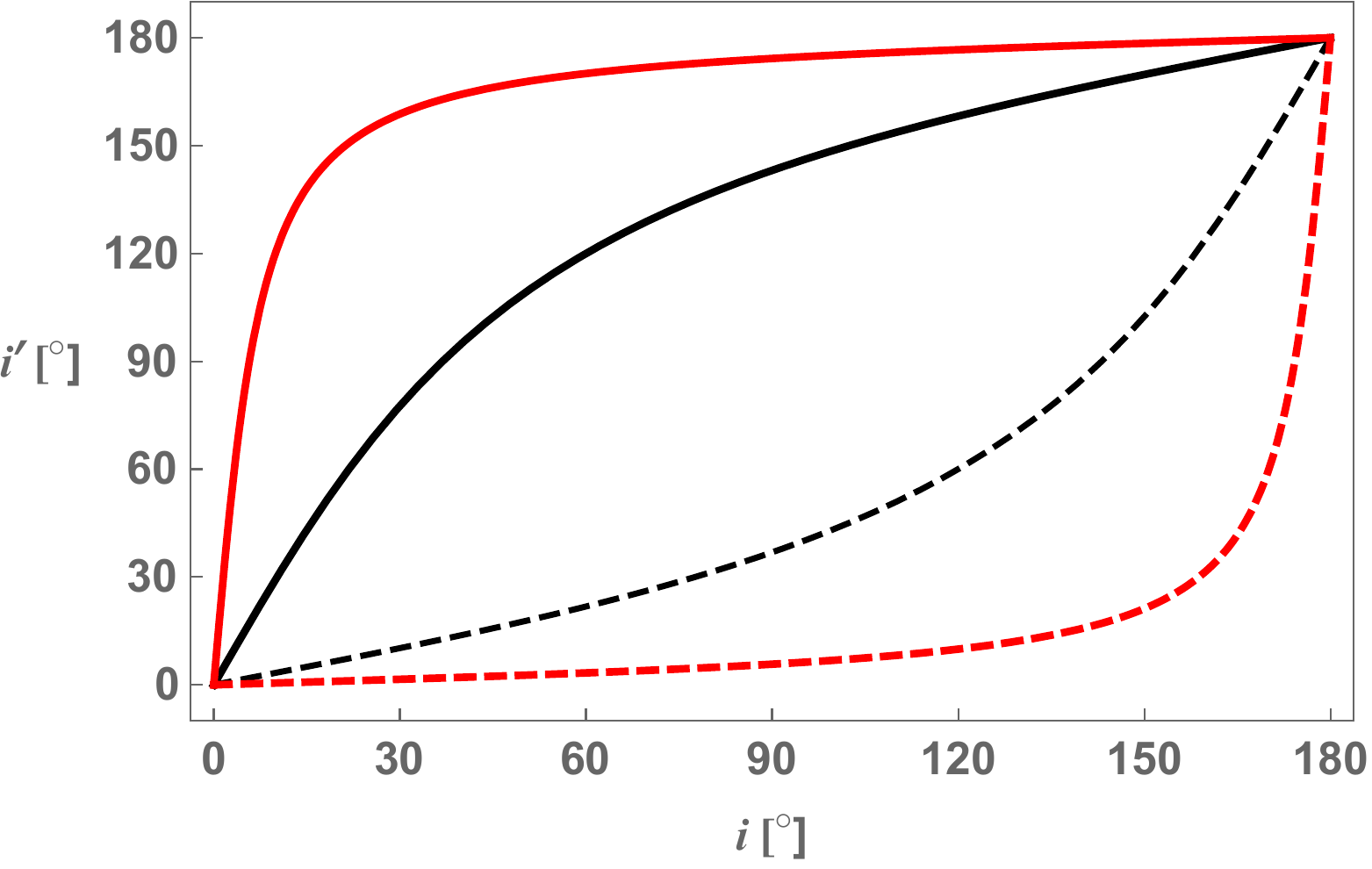}}
\caption{The effect of the special relativistic transformation on the emission angle in the observer frame, $i$. The angles in the comoving frame, $i'$, are shown for the jet and counterjet by the solid and dashed curves, respectively. The top solid and bottom dashed (red) and the two middle (black) curves are for $\Gamma_{\rm j}=10$ and 5/3 ($\beta_{\rm j}=0.8$), respectively. We see that at large $\Gamma_{\rm j}$, the jet-frame viewing angle for most values of $i$ is close to $180\degr$. 
}
\label{transform}
\end{figure}

\begin{figure}
\centerline{\includegraphics[width=7.5cm]{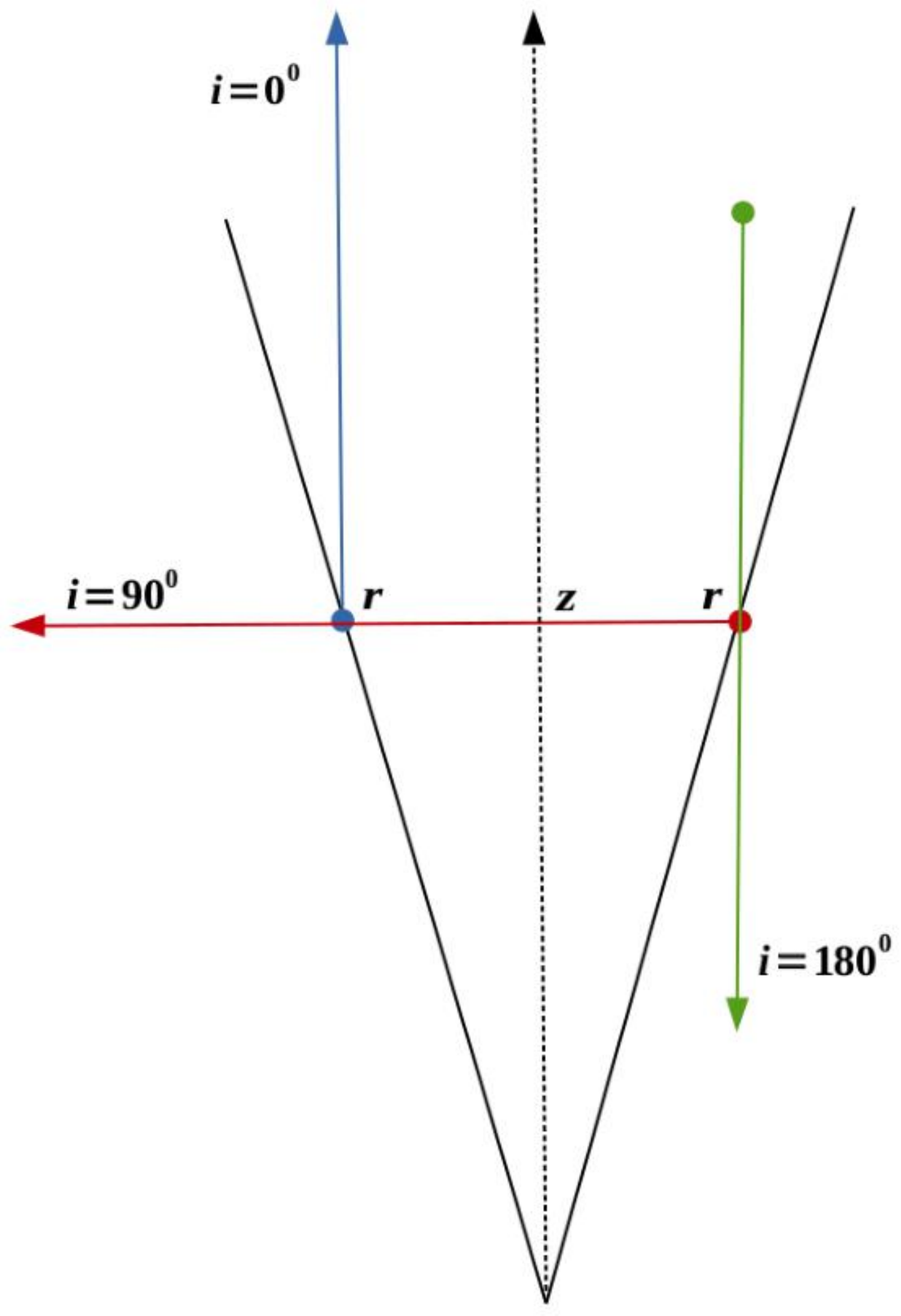}}
\caption{A schematic drawing of the jet and photon paths in the local observer frame. We show three cases of the photon path: (i) along the jet axis and in the direction of the jet motion, in blue, (ii) along the jet axis backward, in green, and (iii) perpendicular to the jet axis, in red. The beginning of the photon paths are marked by the filled circles. The intersections of the photon path with the jet boundary are at the vertical distance $z$ from the origin, and at the radial distance $r$ from the jet axis. 
}
\label{geometry}
\end{figure}

We develop here the formalism of \citet*{zls12}, hereafter ZLS12. We use the notation similar to that of ZLS12, except that now the Doppler factor is $\delta_{\rm j}$, the dimensionless jet length, $\xi$, is in units of the distance of the position of the onset of the synchrotron emission, $z_0$, as in \citet{z14a}, and we take into account the effect of cosmological redshift, which we denote as $z_{\rm r}$. Also, the jet radius is given now accurately as $z\tan\theta_{\rm j}$ rather than by $z\theta_{\rm j}$, where $z$ is the distance from the jet origin and $\theta_{\rm j}$ is the half-opening angle. We consider the model of a conical jet moving with a constant bulk velocity, $\beta_{\rm j}c$, and emitting synchrotron radiation (BK79). 

We are concerned with the angular distribution of the emission of such jets. We observe the jet at an angle, $i$, with respect to the jet axis. This angle in the jet frame is given by
\begin{equation}
\sin i'=\delta_{\rm j}\sin i,\quad \delta_{\rm j}={1\over \Gamma_{\rm j}(1-\beta_{\rm j}\cos i)},
\label{angle}
\end{equation}
where $\Gamma_{\rm j}$ is the bulk Lorentz factor. The counterjet is viewed at $\upi - i$. Examples of the relationship between $i$ and $i'$ are shown in Fig.\ \ref{transform}. We see that at large $\Gamma_{\rm j}$, the jet-frame viewing angle for most values of $i$ is close to $180\degr$. The jet is viewed at $90\degr$ in the jet frame for $\sin i= 1/\Gamma_{\rm j}$, see case (iii) in Fig.\ \ref{geometry}. This case is also equivalent to $\cos i= \beta_{\rm j}$ and $\delta_{\rm j} =\Gamma_{\rm j}$. Hereafter, we omit primes for other quantities in the comoving frame, and use a prime instead for quantities being integrated.

We note, however, that unlike the case of a moving blob of a constant size, often assumed to model blazars, the comoving frame in our case is not stationary. Both the distance from the origin and the radius increase with time in the comoving frame, and the magnetic field and the electron density decrease with time. Thus, a photon directed at $90\degr$ in the comoving frame moves in a changing environment, in spite of its comoving position being constant. Therefore, we perform our calculations in the observer frame, in which the jet is stationary (though its medium is moving). We do it using the emission and absorption coefficients transformed from the comoving frame, see below.

The steady-state electron distribution per unit volume in the comoving frame at a given point (defined in the observer frame) is assumed to be a of a power-law form,
\begin{equation}
N(\gamma) \simeq K\gamma^{-p},\quad p>1,
\label{ngamma}
\end{equation}
where $K$ is the normalization and $\gamma$ is the electron Lorentz factor. Following BK79, we assume conservation of the electron distribution along the jet, and conservation of the energy flux in the toroidal component of the magnetic field. We also define the dimensional energy, $E$, in the observer's frame and a dimensionless one, $\epsilon$, in the jet frame. Thus, we have
\begin{equation}
K=K_0 \xi^{-2},\quad B=B_0\xi^{-1},\quad \xi\equiv z/z_0,\quad \epsilon={(1+z_{\rm r})E\over \delta_{\rm j} m_{\rm e} c^2},
\label{xi_dep}
\end{equation}
where $z_0$ corresponds to the onset of emission.

For calculation of the synchrotron emission and absorption, we assume the magnetic field is tangled \citep{Heinz00}, which implies the emission in the local frame is isotropic. The emission coefficient from isotropic relativistic power-law electrons per unit volume at the jet frame can then be written as,
\begin{equation}
j_{\rm S}(\epsilon,\xi)\simeq
{C_1 \sigma_{\rm T}c K B_{\rm cr}^2  \over 48\upi^2}\left(B\over B_{\rm cr}\right)^{{1+p}\over 2} 
\epsilon^{{1-p}\over 2},
\label{synspecpl}
\end{equation}
where $B_{\rm cr}={m_{\rm e}^2 c^3/e \hbar}$ is the critical magnetic field, $\hbar$ is the reduced Planck constant, $e$ is the electron charge, and $C_1(p)\sim 1$ ($=1$ for $p=3$) follows from averaging over the pitch angle, see, e.g., ZLS12. The emission coefficient in the local observer frame at $E$ equals $\delta_{\rm j}^2 j_{\rm S}(E,\xi)$ (e.g., \citealt{ghisellini00}). The synchrotron self-absorption coefficient averaged over the pitch angle for a power-law electron distribution can be expressed as,
\begin{align}
&\alpha_{\rm S}(\epsilon,\xi)=  {C_2\upi \sigma_{\rm T} K\over  2\alpha_{\rm f}}  \left(B\over B_{\rm cr}\right)^{2+p\over 2} \epsilon^{-{4+p\over 2}}=\alpha_0 \xi^{-{6+p\over 2}}\epsilon^{-{4+p\over 2}},\\
&\alpha_0\equiv {C_2\upi \sigma_{\rm T} K_0\over  2\alpha_{\rm f}} \left(B_0\over B_{\rm cr}\right)^{p+2\over 2},
\label{alphas}
\end{align}
where $\alpha_{\rm f}$ is the fine-structure constant, and the constant $C_2(p)$ ($\simeq 1$ for $p=3$) is given, e.g., in ZLS12. The absorption coefficient in the local observer frame at $E$ equals $\delta_{\rm j}^{-1} \alpha_{\rm S}(E,\xi)$. The source function, $j_{\rm S}/\alpha_{\rm S}$, is then,
\begin{equation}
S(\epsilon,\xi)=S_0 \epsilon^{5/2}\xi^{1/2},\quad S_0\equiv 
\frac{C_1 \alpha_{\rm f} c B_{\rm cr}^{5/2}}{24\upi^3 C_2 B_0^{1/2}}, 
\label{source}
\end{equation}
which becomes $\delta_{\rm j}^3 S(E,\xi)$ in the observer frame.

\section{The jet angular emissivity pattern}
\label{jet}

In order to obtain the spectrum observed from the jet, we need to calculate the emission towards the observer from a given location of the jet in the observer frame and then integrate it over the jet projected area. We can do it by integrating the radiative transfer equation over the line of sight, e.g., equation (1.29) in \citet{rl79}, and then integrate the solution over the projected area. In the conical geometry and for an arbitrary angle, this can be done only numerically, which we do in Appendix \ref{exact}. However, if the viewing angle in the observer frame is $i\sim 90\degr$, see the case (iii) in Fig.\ \ref{geometry}, we can approximate the jet locally as a cylinder, and neglect the variation of $K$ and $B$ along the line of sight. The above approximation also implies that the source function is constant along the line of sight. In this approximation, the radiative transfer equation solves for the observed intensity as $\delta_{\rm j}^3 S [1-\exp(-\tau)]$, where $\tau$ is the optical depth to synchrotron self-absorption integrated over the entire line of sight,
\begin{equation}
\tau(\epsilon,z,x)={2\alpha_{\rm S}(\epsilon,z) \over \delta_{\rm j}\sin i}\left[(z\tan\theta_{\rm j})^2-x^2\right]^{1/2},
\label{tau1}
\end{equation}
where $x$ is the distance perpendicular to both the jet axis and the line of sight. The flux (for either jet or counterjet) is then given by (ZLS12) 
\begin{equation}
{{\rm d}F\over {\rm d}E}={(1+z_{\rm r})\delta_{\rm j}^3 \sin i\over m_{\rm e} c^2 D^2} \int_{z_0}^\infty {\rm d}z\, S\int_{-z\tan\Theta_{\rm j}}^{z\tan\Theta_{\rm j}}{\rm d}x\left[1-\exp(-\tau)\right],
\label{int1}
\end{equation}
where $D$ is the luminosity distance. Since uppermost parts of the jet usually contribute weakly to the total flux, we assume here the jet extends to infinity. Hereafter, the redshift terms are included, generalizing the corresponding equations in ZLS12. The unit of energy in the flux is assumed to be the same as the unit of photon energy (e.g., both can be in eV or erg), which results in the unit of the flux of cm$^{-2}$ s$^{-1}$. To get the flux in, e.g., erg and the photon energy in eV, we need to multiply the above formula by erg/eV.

Above the synchrotron break energy, $E_{\rm t}$, see equation (\ref{break1}) below, the entire jet emission is optically thin, with $\tau<1$. Below it, the jet is optically thick from $z_0$ up to some distance, and optically thin above it. This results in the spectral index of the partially optically thick spectrum of $\alpha=0$ regardless of $p$ (as obtained by BK79). In order to get the normalization of this spectrum, we can solve equation (\ref{int1}). This is equivalent to substituting equation (23) in equation (22) of ZLS12, which yields
\begin{equation}\begin{aligned}
\frac{{\rm d}F}{{\rm d}E}= &\frac{C_1 C_3 (1+z_{\rm r})\alpha_{\rm f}\delta_{\rm j}^3(B_0 z_0)^2 \sin i\tan\theta_{\rm j}}{24 \upi^3 C_2 m_{\rm e}c D^2}\times \\
&\left(\upi C_2 \sigma_{\rm T} K_0 z_0 B_{\rm cr}\tan\theta_{\rm j}\over \alpha_{\rm f} \delta_{\rm j}B_0\sin i\right)^{5\over 4+p}\!\!\!,
\label{thick}
\end{aligned}\end{equation}
which is valid for $E\ll E_{\rm t}$. $C_3(p)$ is given by a dimensionless double integral defined in ZLS12, which we integrate here analytically,
\begin{equation}\begin{aligned}
C_3(p)&\equiv \int_0^\infty {\rm d}\zeta\,\zeta^{3\over 2}  \int_{-1}^1 {\rm d}\psi \left\{1-\exp\left[-\zeta^{-{p+4\over 2}}\left(1-\psi^2\right)^{1\over 2}\right]\right\}\\
&={2\sqrt{\upi}\, \Gamma\left(5\over 2 p+8\right) \Gamma\left(p-1\over p+4\right)\over (p+9) \Gamma\left(p+9\over 2 p+8\right)},
\end{aligned}\end{equation}
where $\Gamma$ is the Gamma function. Example values of $C_3$ are $\simeq 3.61$, 2.10, 1.61 for $p=2$, 3, 4, respectively. The flux dependence on the angle is $\propto \delta_{\rm j}^{(7+3 p)/(4+p)}(\sin i)^{(p-1)/(4+p)}$, see also \citet{cawthorne91}. This can be compared to the angular dependence of a steady-state optically-thin jet of $\propto \delta_{\rm j}^{2-\alpha}$ (BK79; \citealt{lb85,sikora97}) for $\alpha=0$. We see that at large angles the two dependencies are relatively similar, having similar powers of $\delta_{\rm j}$, e.g., 13/6 and 16/7 for the partially optically-thick jet with $p=2$, 3, respectively, vs.\ 2 for an optically thin jet. However, at small viewing angles we have $\sin i\ll 1$, which leads to a strong reduction of the flux, formally to null at $i=0$.

However, we point out that the approximation of equation (\ref{int1}) breaks down when the viewing angle is substantially different from $90\degr$, in particular for $i\sim 0$ or  $i\sim 180\degr$, see Fig.\ \ref{geometry}. This is because a given line of sight goes through regions with variable magnetic field and the electron normalization, and $S$ is no longer constant along it. Furthermore, the length along a line of sight is no longer given by that in a cylinder, which is assumed in equation (\ref{tau1}). Generally, the integration of the attenuated source function over the jet volume needs to be done numerically, see Appendix \ref{exact}. Furthermore, the treatment needs to be changed for $i<\theta_{\rm j}$, in which case we no longer see the jet side, but instead view it from the top. An analogous effect occurs for the bottom view. These cases require a change of the method of the integration of the radiative transfer solution over the projected area (Appendix \ref{exact}). 

Thus, the question arises whether the angular dependence dominated by the above power of $\sin i$ actually takes place. In order to address this question, we consider the limiting approximation in which we view the jet at $i=0$, see case (i) in Fig.\ \ref{geometry}. A conical jet has divergent velocities, see, e.g., \citet*{moderski03}, which effect we neglect, and assume that $i=0$ along all the lines of sight through the jet. We note that the entire emission originating at a given distance, $z$, is now attenuated by the same optical depth, corresponding to the positions $>z$. Thus, we can simplify the radiative transfer equation for this case and multiply the source function by the jet cross section, $\upi r^2 S(r)$, where $r=z\tan\theta_{\rm j}$ is the jet radius. We can then write the radiative transfer equation directly for the flux, $\propto \int r(\tau)^2 S(\tau)\exp(-\tau){\rm d}\tau$, where $\tau(r)$ is measured from the top of the jet down to the intersection of the line of sight with the jet boundary (at $z=r/\tan\theta_{\rm j}$). Uppermost parts of the jet contribute weakly to the total emission, as well as the part closest to the jet origin are strongly self-absorbed, which is the case for photon energies well below the synchrotron break, $E_{\rm t}$, see below. Thus we calculate the above integral from 0 to $\infty$.

\begin{figure}
\centerline{\includegraphics[width=\columnwidth]{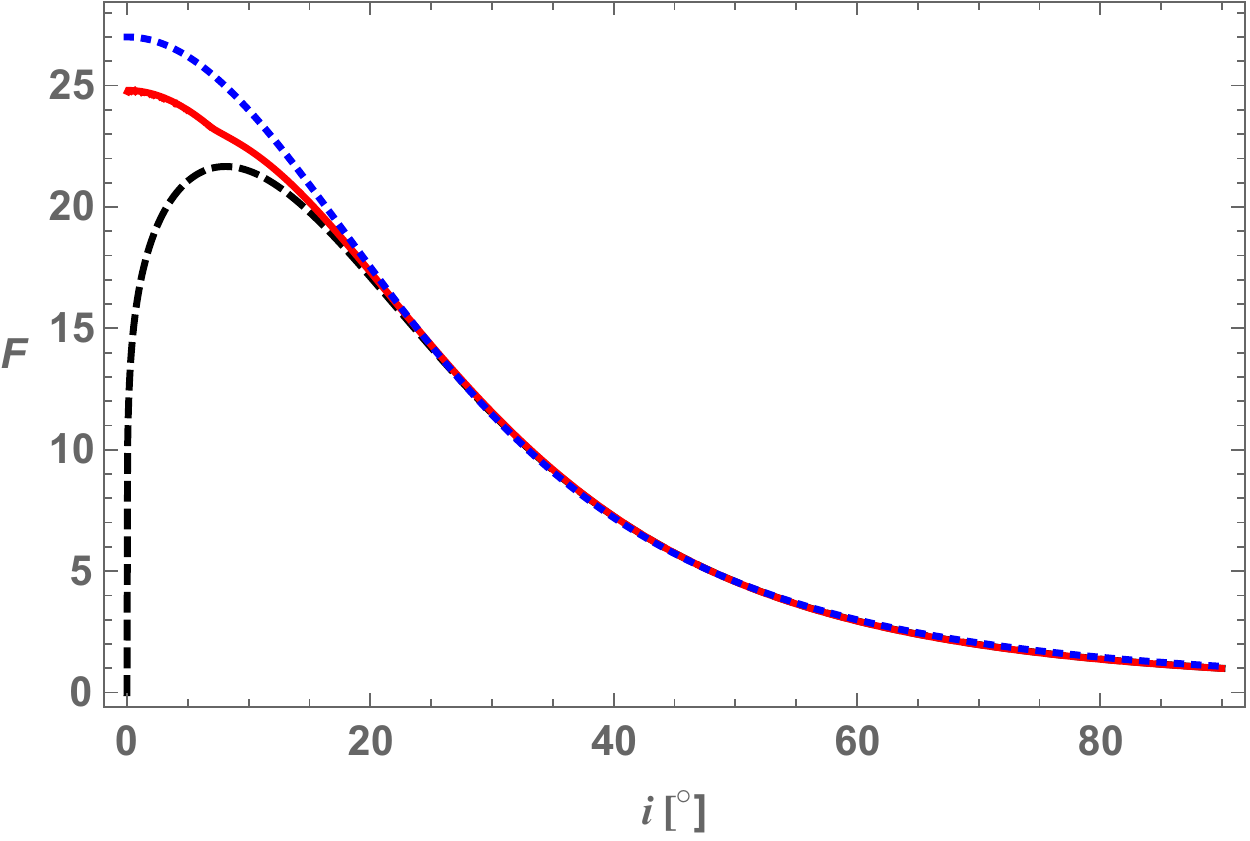}}
\caption{An comparison of the jet flux angular distributions, at $\beta_{\rm j}=0.8$ ($\Gamma_{\rm j}=5/3$). The blue dotted curve shows the distributions for an optically thin steady-state jet, i.e., $F\propto \delta_{\rm j}^2$. The red solid curve shows the exact dependence for a partially optically-thick jet, see Appendix \ref{exact}, for $p=2$ and $a=0.2$. This dependence reaches the value of the small-angle limit of equation (\ref{onaxis}) at $i=0$. The black dashed curve shows the cylindrical approximation of equation (\ref{thick}), which is exact at $90\degr$, but it breaks down at small angles. The flux is normalized to unity at $i=90\degr$.
}
\label{angular}
\end{figure}

We first calculate $\tau$ from a given $h$ up to infinity at $i\simeq 0$, 
\begin{equation}
\tau(\xi)=\delta_{\rm j0}^{-1} z_0\!\!\int_{\xi}^\infty\!\! {\rm d}\xi' \alpha_{\rm S}(\epsilon,\xi')=\tau_0 \epsilon^{-{4+p\over 2}}\xi^{-{4+p\over 2}}\!, \,\,\, \tau_0\equiv {2 \alpha_0 z_0\over (4+p)\delta_{\rm j0}}
,\label{taus}
\end{equation}
where $\delta_{\rm j0}=\sqrt{(1+\beta_{\rm j})/(1-\beta_{\rm j})}$ is the Doppler factor at $i=0$. We note that equation (\ref{taus}) is equivalent to the limit of $i\rightarrow 0$ of the expression for $\tau$ in the side-view case, see, e.g., equation (19) of ZLS12. Namely, in that case, ${\rm d}\tau=\alpha_{\rm S}{\rm d}x/(\delta_{\rm j}\sin i)$. We have $\sin i={\rm d}x/({\rm d}x^2+{\rm d}z^2)^{1/2}$. Thus, ${\rm d}\tau=\alpha_{\rm S}({\rm d}x^2+{\rm d}z^2)^{1/2}/\delta_{\rm j}$, which, for $i=0$, equals ${\rm d}\tau=\alpha_{\rm S}{\rm d}z/\delta_{\rm j0}$, as above. We also note that although $\tau$ is invariant between different frames, it does depend on $\delta_{\rm j}$.

We can then write the equation for the flux as
\begin{equation}
{{\rm d}F\over {\rm d}E}={\upi (1+z_{\rm r})\delta_{\rm j0}^3 (z_0 \tan\theta_{\rm j})^2\over m_{\rm e} c^2 D^2} \int_0^\infty {\rm d}\tau\, \xi(\tau)^2 S(\tau) \exp(-\tau),
\label{Fi0}
\end{equation}
where $\xi(\tau)$ follows from equation (\ref{taus}) and $S(\tau)=S[\xi(\tau)]$. The above integral yields a Gamma function, and we obtain a spectrum with $\alpha=0$,
\begin{equation}
{{\rm d}F\over {\rm d}E}=\frac{C_1 (1+z_{\rm r})\alpha_{\rm f} \delta_{\rm j0}^3(B_0 z_0\tan\theta_{\rm j})^2}{24 \upi^2 C_2 m_{\rm e}c D^2}\left[\upi C_2 \sigma_{\rm T} K_0 z_0 B_{\rm cr}\over (4+p)\alpha_{\rm f}\delta_{\rm j0} B_0\right]^{5\over 4+p}\!\! \Gamma\left(\frac{p-1}{p+4}\right).
\label{onaxis}
\end{equation}
The flux dependence on the Doppler factor is the same as of the side-view spectrum, equation (\ref{thick}), namely $\propto \delta_{\rm j0}^{(7+3 p)/(4+p)}$. Parenthetically, we note that the often-used approximation of calculating the emission from the part with $\tau\leq 1$ as optically thin and neglecting the contribution from the optically-thick part yields almost the same result, with the Gamma function above replaced by inversion of its argument, $(p+4)/(p-1)$, which fractional error is less than 13 per cent for $p\leq 4$. 

It is of interest to calculate the ratio of the on-axis emission, equation (\ref{onaxis}), to that seen from a side, equation (\ref{thick}),
\begin{equation}
{{\rm d}F/{\rm d}E(i\simeq 0)\over {\rm d}F/{\rm d}E(i\sim 90\degr)}=
{\upi \delta_{\rm j0}^3\tan\theta_{\rm j}\over C_3\delta_{\rm j}^3\sin i} \left[\delta_{\rm j}\sin i\over (4+p)\delta_{\rm j0}\tan\theta_{\rm j}\right]^{5\over 4+p}\!\!\Gamma\left(\frac{p-1}{p+4}\right).
\label{ratio}
\end{equation}
We see that, apart from the dependence on the viewing angle, this ratio depends only on the jet opening angle and velocity, and on the electron index, $p$.

In order to get a characteristic value of this ratio, we calculate it at $\sin i\!=\!1/\Gamma_{\rm j}$, $i'=90\degr$. For that angle, $\delta_{\rm j0}/\delta_{\rm j}\!=\!1+\beta_{\rm j}$. Then, we express the jet opening angle as a fraction of $1/\Gamma_{\rm j}$, $\tan\theta_{\rm j}\!=\!a/\Gamma_{\rm j}$. In theoretical models, often $a\!=\!1$ is assumed (e.g., \citealt{zamaninasab14}), but observationally $a\!<\!1$ is found. In particular, \citet{pushkarev09} and \citet{clausen13} found the average values in their samples of extragalactic jets of $a\!\simeq\! 0.13$, $\simeq\! 0.2$, respectively. For jets in black-hole binaries, \citet*{millerjones06} also found $a\!\ll\! 1$. We note that the factor $a$ can be theoretically connected in some jet models to the magnetization parameter, $\sigma\!\sim\! a^2$ \citep*{tmn09,komissarov09}. 

Using the above relations, we obtain
\begin{equation}
{{\rm d}F/{\rm d}E(i\simeq 0)\over {\rm d}F/{\rm d}E(i'=90\degr)}=
{\upi (1+\beta_{\rm j})^{7+3 p\over p+4} \over C_3(p)(p+4)^{5\over p+4}} a^{p-1\over 4+p}\Gamma\left(\frac{p-1}{p+4}\right).
\label{ratio0}
\end{equation}
We see that this ratio depends only on the fractional jet opening angle, $a$, the electron index, and $1+\beta_{\rm j}$, which factor can assume values only between 1 and 2. We also see that this ratio is large unless $a$ is very small. For example, for $\beta_{\rm j}\simeq 1$ and $p=2$, 3, 4, it is $\simeq 4.88 a^{1/6}$, $5.73 a^{2/7}$, $6.54 a^{3/7}$, respectively. 

The solid (red) curve in Fig.\ \ref{angular} shows an example of the exact angular distributions of a partially optically-thick jet, calculated in Appendix \ref{exact}, which reaches the values of the analytical formuale at $i=0\degr$ and $180\degr$. We see that the flux seen on axis is substantially higher than the maximum of the flux in the large-angle approximation, shown by the dashed (black) curve. This appears to be a new result; e.g., \citet{cawthorne91} derived the maximum flux and the corresponding viewing angle using the large-angle approximation (as for the dashed curve in Fig.\ \ref{angular}), while we find the overall maximum to occur at $i=0$, with the flux decrease with the decreasing $i$ of that approximation being spurious. 

We then compare the values of the synchrotron break energy, $E_{\rm t}$. For the top view, we solve $\tau(1)=1$ with equation (\ref{taus}),
\begin{equation}
E_{\rm t}(i\simeq 0)={\delta_{\rm j0} m_{\rm e}c^2\over 1+z_{\rm r}}\left[{2\alpha_0 z_0\over (4+p)\delta_{\rm j0}}\right]^{2\over 4+p}.
\label{break0}
\end{equation}
In the side-view case, we have (ZLS12),
\begin{equation}
E_{\rm t}(i\sim 90\degr)={\delta_{\rm j} m_{\rm e}c^2\over 1+z_{\rm r}}\left[2\alpha_0 z_0\tan \theta_{\rm j}\over \delta_{\rm j}\sin i\right]^{2\over 4+p}.
\label{break1}
\end{equation}
With the same assumptions as above, we obtain with equation (\ref{break0}),
\begin{equation}
{E_{\rm t}(i\simeq 0)\over E_{\rm t}(i'=90\degr)}=(1+\beta_{\rm j})^{2+p\over 4+p}
\left[(4+p)a\right]^{-{2\over 4+p}}.
\label{eratio}
\end{equation}
This ratio is typically $\sim$1, e.g., it is $\simeq$1.5 for $\beta_{\rm j}=1$, $p=2$, $a=0.2$.

\section{The counterjet and large angle emission}
\label{cjet}

The emission of the counterjet corresponds to the viewing angle of $\upi - i$. We note that $\sin (\upi - i)=\sin i$, and $\cos (\upi - i)=-\cos i$. The Doppler factor of the counterjet is then
\begin{equation}
\delta_{\rm cj}={1\over \Gamma_{\rm j}(1+\beta_{\rm j}\cos i)}.
\label{dcjet}
\end{equation}
Equation (\ref{thick}) then applies to to the counterjet emission with the substitution of $\delta_{\rm j}\rightarrow \delta_{cj}$. When we consider the flux ratio between the jet and counterjet in the side-view approximation, the angular dependencies on $\sin i$ in equation (\ref{thick}) cancel each other. 

Thus, we find the jet to counterjet flux ratio under the assumption of both components being intrinsically symmetric of
\begin{equation}
R\equiv {{\rm d}F_{\rm j}/{\rm d}E\over {\rm d}F_{\rm cj}/{\rm d}E} = \left(\delta_{\rm j}\over \delta_{\rm cj}\right)^{7+3 p\over 4+p}, \quad {\delta_{\rm j}\over \delta_{\rm cj}}={1+\beta_{\rm j}\cos i\over 1-\beta_{\rm j}\cos i}.
\label{fratio}
\end{equation}
The power-law index of this dependence changes from 2 for $p=1$ to $\simeq$3 for $p\gg 1$, different from the index of 2 for an optically thin jet with $\alpha=0$. The total emission is, obviously, the sum of the fluxes from the jet and the counterjet. We note that equation (22) of ZLS12 implies an incorrect ratio, due to the neglect of the different value of $E_{\rm t}$ between the jet and counterjet. Equation (\ref{fratio}) can be solved for $\beta_{\rm j}$,
\begin{equation}
\beta_{\rm j}={1\over \cos i}{R^{4+p\over 7+3 p}-1\over R^{4+p\over 7+3 p}+1}.
\label{beta}
\end{equation}

We now consider the case of the counterjet emission at $i\sim 0$. At this approximation, we have considered the emission from the top of the jet, equations (\ref{Fi0}--\ref{onaxis}). However, this counterjet emission corresponds to that from the jet bottom, $i\simeq 180\degr$, see the case (ii) in Fig.\ \ref{geometry}, for which those equations are not applicable. We can still use the approximation of a ray along the jet axis, but the direction of the emission is opposite. This also means that we cannot multiply the source function by the jet cross section, since this emission will now be attenuated by different $\tau$ depending on the distance from the axis. Using radiative transfer, we write
\begin{equation}
{{\rm d}F\over {\rm d}E}={2\upi (1+z_{\rm r})\delta_{\rm cj0}^3 \over m_{\rm e} c^2 D^2} \int_0^\infty {\rm d}r\, r \int_0^{\tau(r)} {\rm d}\tau'\, S(\tau') \exp(\tau'-\tau),
\label{Fcj}
\end{equation}
where $\tau(r)$, given by equation (\ref{taus}), corresponds to the intersection of the line of sight with the jet boundary at the radial distance $r$. Then the optical depth along the ray emitted at $i=180\degr$ is $\tau-\tau'$, i.e., it is measured from the intersection with the boundary. We can now change the variable of the outer integration to $\tau$ [using equation (\ref{taus})]. The resulting double integral can be calculated analytically, and the final result is
\begin{equation}
\begin{aligned}
{{\rm d}F\over {\rm d}E}&=\frac{C_1 (1+z_{\rm r})\alpha_{\rm f} \delta_{\rm cj0}^3(B_0 z_0\tan\theta_{\rm j})^2}{6 \upi C_2(4+p) m_{\rm e}c D^2}\times\\
& \left[\upi C_2 \sigma_{\rm T} K_0 z_0 B_{\rm cr}\over (4+p)\alpha_{\rm f}\delta_{\rm cj0} B_0\right]^{5\over 4+p}\left(\cot{5\upi\over 4+p} - \cot{\upi\over 4+p}\right){\Gamma\left(\frac{-4}{p+4}\right)\over
 \Gamma\left(\frac{1}{p+4}\right)}.\label{flux_cj}
\end{aligned}\end{equation}
We can obtain the jet-to-counterjet flux ratio for this case using equation (\ref{onaxis}). We find it is almost the same, within $\sim$10 per cent, as that given by equation (\ref{fratio}). This shows that equation (\ref{fratio}) is valid quite generally, in spite of the underlying approximation breaking down at $i$ substantially different from $90\degr$. Naturally, equation (\ref{flux_cj}) with a replacement of $\delta_{\rm cj0}\rightarrow \delta_{\rm j}$ also gives the jet emission at $i\simeq 180\degr$.

We also note that we have neglected here a possible (and likely in a range of angles) obscuration of the counterjet emission at $i\sim 0$, by the accretion disc and stellar wind. Also, the jet will synchrotron-absorbed reprocess the counterjet synchrotron emission.

\section{Application to Cyg X-1}
\label{cygx1}

\begin{figure}
\centerline{\includegraphics[width=\columnwidth]{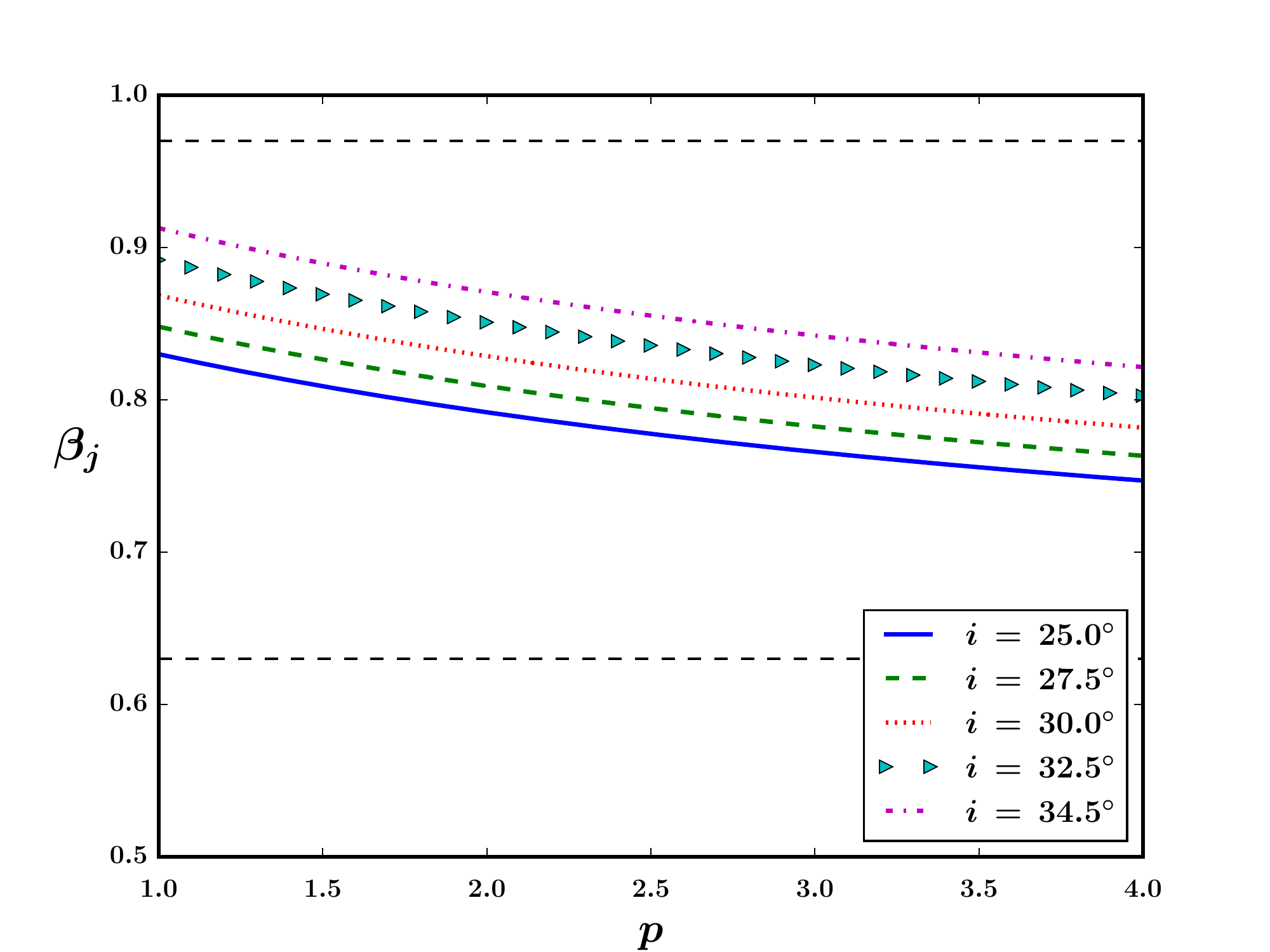}}
\caption{The lower limit on the velocity of the jet from Cyg X-1 as a function of the viewing angle, $i$, and the electron index, $p$. We use the range of the viewing angles found by \citet{ziolkowski14} and the lower limit on the ratio of the jet-to-counterjet flux ratio obtained by \citet{stirling01} of $R\geq 50$. The horizontal lines correspond to the previous limits given in their table 2.
}
\label{beta_p}
\end{figure}

\citet{stirling01} have obtained radio maps of the black-hole binary Cyg X-1 using VLBA and VLA at 8.4 GHz. They have found no evidence for the presence of a counterjet, and constrained the flux ratio to $R\gtrsim 50$. They assumed $i=40\degr$ and optically-thin emission in calculating constraints on the jet velocity. Currently, a lower value of $i$ appears to be the inclination of the orbit of Cyg X-1, in particular \citet{orosz11} found $i\simeq 27\pm 1\degr$. On the other hand, \citet{ziolkowski14}, considering also the evolutionary status of the system, found $i\simeq 29^{+5}_{-4}\degr$ as the most likely range. For this range, equation (\ref{beta}) at $p=2$ gives $\beta_{\rm j}\geq 0.82_{-0.03}^{+0.05}$, corresponding to $\Gamma_{\rm j}\geq 1.75_{-0.11}^{+0.25}$. The value of $p$ has been constrained by broad-band spectral models of ZLS12, \citet*{mzc13} and \citet{z14b}, which give the most likely range of $1.4\lesssim p \lesssim 2.5$. The resulting limits on $\beta_{\rm j}$ are relatively insensitive to the value of $p$, as illustrated in Fig.\ \ref{beta_p}. A caveat for this result is that the jet is likely aligned with the black-hole rotation axis, which may be misaligned with the normal to the binary plane. In this case, the jet inclination may be not given by the above constraints. Possible indications for this to be the case are given by the results of \citet{tomsick14} and \citet{walton16}, who have fitted X-ray data from observations of Cyg X-1 in the soft spectral state (in which the inner disc is expected to extend to the innermost stable orbit) by \nustar\/ and \suzaku, and \nustar, respectively. Their fits gave the inclination of the inner disc substantially larger than $30\degr$, namely $\simeq\! 70\degr$ and $\simeq\! 40\degr$, respectively, which would indicate the plane of the inner disc inclined with respect to the orbital axis. We caution, however, that this may be due to the limitations of the used models. \citet{tomsick14} used the Compton reflection model of \citet{ross05}, which averages over all viewing angles. Then, \citet{walton16} used the model of \citet{garcia10}, which calculates angle-dependent reflection, but both models convolve their reflection spectra using the relativistic code of \citet{dauser10}, which neglects a number of important effects, as discussed in \citet*{niedzwiecki16}. Thus, we consider the issue of the misalignment to be open.

Other constraints on the jet velocity in Cyg X-1 are by \citet{gleissner04}, who claimed $\beta_{\rm j}\lesssim 0.7$ or so based on non-detection of short-time scale correlations between radio and X-ray emission, though this limit appears model-dependent. Then \citet*{mbf09} estimated $0.3\lesssim \beta_{\rm j}\lesssim 0.8$, which also relies on a number of assumptions. Still, if we accept those constraints at face value, the most likely jet velocity in Cyg X-1 becomes $\beta_{\rm j}\simeq 0.8$.

\citet{stirling01} also constrained the half-opening jet angle of the projection of the jet on the sky as $\lesssim 2\degr$. We note here that the actual half-opening angle is the one after de-projection, i.e., multiplied by $\sin i$ (e.g., \citealt{konigl81}). Given that $i\simeq 30\degr$, $\theta_{\rm j}\lesssim 1\degr$. Since the lower limit on $\Gamma_{\rm j}$ is rather low, see above, this upper limit on the half-opening angle implies a very small factor $a\equiv \Gamma_{\rm j}\tan\theta_{\rm j}$, unless $\Gamma_{\rm j}$ is much higher than the lower limit of \citet{stirling01} obtained from the absence of an observed counterjet. If $\beta_{\rm j}=0.8$ ($\Gamma_{\rm j}=5/3$), $a\lesssim 0.03$. This implies an extremely efficient collimation mechanism, e.g., that related to a very low jet magnetization in the models of \citet{tmn09} and \citet{komissarov09}. Also, this coefficient is substantially lower than those typically seen in blazars, where $a\simeq$0.2 \citep{clausen13}.

\section{Conclusions}

We have studied the extended synchrotron jet model, originally proposed by BK79. We have considered three limiting analytical approximations to the flux vs.\ the viewing angle. In one, usually assumed, the jet is viewed sideways in the comoving frame. This approximation implies that the flux becomes null when the jet is viewed on axis, with $i\simeq 0$, e.g., \citet{cawthorne91}. However, we point out that the above approximation breaks down in the low-$i$ regime, since the jet is no longer viewed sideways in the comoving frame. We have considered another limiting case, of the jet viewed on axis. We have found an analytical solution of the radiative transfer integrated over the jet cross section in that case. We have found out that this emission is rather strong, corresponding to the global maximum of the flux as a function of the viewing angle. We have also calculated the emission corresponding to the emission angle of $i\sim 180\degr$, which also corresponds to the counterjet emission in the case of $i\simeq 0$. Then, we have obtained the general exact solution at an arbitary viewing angle numerically, described in Appendix \ref{exact}.

We have applied our results to the black-hole binary Cyg X-1. Given the jet-to-counterjet flux ratio of $\gtrsim$50 found observationally \citep{stirling01} and the current estimate of the inclination of $i\simeq 29^{+5}_{-4}\degr$ \citep{orosz11,ziolkowski14}, we have found $\beta_{\rm j}\gtrsim 0.8$, $\Gamma_{\rm j}\gtrsim 1.6$. Combining it with other published constraints, the most likely value is $\beta_{\rm j}\simeq 0.8$. We have also pointed out that when the projection effect is taken into account, the radio observations imply the jet half-opening angle of $\theta_{\rm j}\lesssim 1\degr$, a half of the value given by \citet{stirling01}. If $\Gamma_{\rm j}$ is not much above the counterjet limit, the opening angle is $\theta_{\rm j}\ll 1/\Gamma_{\rm j}$, and much lower than the values typically observed in blazars.

\section*{ACKNOWLEDGMENTS}
We thank Marek Sikora for valuable discussions and the referee for valuable comments and suggestions. This research has been supported in part by the Polish National Science Centre grants 2012/04/M/ST9/00780, 2013/10/M/ST9/00729 and 2015/18/A/ST9/00746.

\onecolumn
\appendix
\section{Exact calculation of the angular dependence}
\label{exact}

As stated in Section \ref{jet}, we first obtain the intensity emitted by a given point of the jet projection in a given direction, and then integrate it over the projected area. In order to efficiently deal with the geometry of an inclined cone and its projection, we set up two coordinate systems in the observer frame. One is the usual jet coordinate system, with the $z$ axis along the jet axis, and the $x$ and $y$ axes orthogonal to it. The other system has the origin at the top of the jet, at the height $Z$, at $(0,0,Z)$ in the jet coordinates, and it is rotated clockwise by the viewing angle, $i$. This corresponds to an anticlockwise rotation of the jet itself. Since the optically-thin jet emission declines fast with the distance, the results are almost independent of the assumed value of $Z$ as long as it is much larger than the height at which the emission at a given frequency becomes optically thin. The coordinates in the rotated system are marked by the subscript $i$. The transformations from the jet system to the rotated one and the reverse one are
\begin{equation}
\begin{bmatrix}
    x_{\rm i}\\
    y_{\rm i}\\
    z_i\\
  \end{bmatrix}
  =R_{\rm x}\left(\begin{bmatrix} x\\ y\\ z\\ \end{bmatrix}-\begin{bmatrix} 0\\ 0\\ Z\\ \end{bmatrix}\right)
  =\begin{bmatrix} 
  1 & 0 & 0\\
  0 & \cos i & -\sin i\\
  0 & \sin i & \cos i\\
  \end{bmatrix}
  \begin{bmatrix}
  x\\
  y\\
  z-Z\\
  \end{bmatrix}
  =\begin{bmatrix}
  x\\
  y \cos i + (Z-z) \sin i\\
  y \sin i + (z-Z) \cos i\\
  \end{bmatrix},\quad {\rm and}\quad
\begin{bmatrix}
    x\\
    y\\
    z\\
  \end{bmatrix}
  =\begin{bmatrix}
  x_i\\
  y_i \cos i + z_i \sin i\\
  -y_i \sin i + z_i \cos i +Z\\
  \end{bmatrix},
  \label{rotated}
\end{equation}
respectively, where $R_{\rm x}$ is the three-dimensional clockwise rotation matrix fixing the $x$ axis.

The way we calculate the intensity depends on the viewing direction. For viewing from either the top or the side, $i < \upi - \theta_{\rm j}$, $\tau$ is taken to be 0 when the ray leaves the jet, and integrated backward to its maximum value at the ray origin. For viewing the jet from the back, we follow the usual convention of the radiative transfer, and take $\tau$ to be 0 at the origin and integrate it to its maximum value at which the ray leaves the jet. This corresponds to the expressions for the intensity in the jet frame (see Section \ref{def}) of
\begin{equation}
I=\begin{cases}
\delta_{\rm j}^3\int_0^{\tau } {\rm e}^{-\tau '} S\left(\tau '\right) \, {\rm d}\tau', &i \in (0,\upi-\theta_{\rm j});\cr
\delta_{\rm j}^3\int_0^{\tau } {\rm e}^{\tau'-\tau} S\left(\tau '\right) \, {\rm d}\tau', &i \in (\upi-\theta_{\rm j},\theta_{\rm j}),\cr
\end{cases}\quad 
{\rm d}\tau'=\frac{{\rm d}l' \alpha \left(l'\right)}{\delta_{\rm j} },
\label{intensity}
\end{equation}
where $l'$ is the running length measured along the ray. We then change the variable of integration from $l'$ to the height, $z$, measured in the jet coordinates as it is the only geometric quantity that the source and absorption functions vary with, the other quantities being constant. This gives us
\begin{align}
&I=
\begin{cases}
	\delta_{\rm j}^2 (\cos i)^{-1} \int_{z_{2}}^{z_{1}} j\left(z\right) {\rm e}^{-\tau \left(z\right)} {\rm d}z, &i \in (0,\upi-\theta_{\rm j}); 
	\cr
\delta_{\rm j}^2 [\cos (i-\upi)]^{-1} \int_{z_{1}}^{z_{2}} j\left(z\right) {\rm e}^{\tau(z)-\tau \left(z_{1}\right)} {\rm d}z, &i \in (\upi-\theta_{\rm j},\upi),\cr
	\end{cases}\\
   &\tau(z)=\frac{2 \alpha_0 z_0^{\frac{p+6}{2}} \epsilon ^{-\frac{p+4}{2}}
   }{(p+4) \delta_{\rm j} }\left(z^{-\frac{p+4}{2}}-z_{2}^{-\frac{p+4}{2}}\right)\times
	\begin{cases}(\cos i)^{-1},
		  &i \in (0,\upi-\theta_{\rm j}); \cr
[\cos (i-\upi)]^{-1}, &i \in (\upi-\theta_{\rm j},\upi),\cr
	\end{cases}\label{tau}
\end{align}
where the boundaries $z_1$ and $z_2$ are derived below. Since we are concerned with the emission in the partially optically-thick regime, we calculate the emission of the entire jet, down to $z=0$, as in Sections \ref{jet} and \ref{cjet}, and do not impose in numerical calculations $z_{1,2}\geq z_0$. From equation (\ref{tau}), we can readily determine the value of $z$ corresponding to $\tau=1$. 

Given that we calculate the intensity in the rotated coordinate system, we have to find the heights of the intersections as functions of $x_i$ and $y_i$. For a given a point, $(x,y,z)$, in the jet coordinates and the viewing angle, $i$, we can define a straight line at that angle that passes through the point. We need to find the intersections of this line with the cone, given by 
\begin{equation}
z=\cot\theta_{\rm j}\sqrt{x^2+y^2}.
\end{equation}
The intersections with the cone can be found by squaring both sides and solving the resulting quadratic equation. However, squaring the equation for the cone also gives us a second cone flipped over the $x$-$y$ plane (the counterjet). When the viewing angle is less than the jet opening angle, the first intersection is on the second cone and when it is greater than $\upi-\theta_{\rm j}$, the second intersection will be on the second cone, so the functions are assigned the value of infinity at these angles so that these cases can be separated. The equations are then written in terms of the rotated coordinate system using the transformation of equation (\ref{rotated}) with $z_i=0$, since the area will be projected onto the $x_i$-$y_i$ plane. This yields,
\begin{align}
&y_{1}(x_i,y_i,i) =
\begin{cases}
		\displaystyle{-\frac{\omega+\cot i (Z-y_i \sin i)-y_i
   \cos i \cot ^2 i}{\cot ^2i-\cot ^2 \theta_{\rm j}}},  & i \in (\theta_{\rm j},\upi); \cr
		\infty & i \in (0,\theta_{\rm j}),\cr
	\end{cases}\\
&y_{2}(x_i,y_i,i) =\begin{cases}
		\displaystyle{\frac{\omega-\cot i (Z-y_i \sin i)+
   y_i\cos i \cot ^2i}{\cot ^2i-\cot ^2\theta_{\rm j}}},  & i \in (0,\upi-\theta_{\rm j}), \cr
		\infty &i \in (\upi-\theta_{\rm j},\upi),\cr
	\end{cases}\\
  &\omega=\cot \theta_{\rm j} \sqrt{ (Z-y_i \sin i)^2-2 y_i
   (Z-y_i \sin i)\cos i \cot i +\left(y_i^2 \cos
   ^2 i+x_i^2\right)\cot
   ^2 i -x_i^2 \cot
   ^2\theta_{\rm j}}.
\end{align}
The heights of the intersections in the jet coordinates are then given by
\begin{align}
&z_{1}(x_i,y_i,i) = \min \left(\cot \theta_{\rm j}\sqrt{y_{1}^2 + x_i^2}, Z\right),\\
&z_{2}(x_i,y_i,i) = \min \left(\cot \theta_{\rm j}\sqrt{y_{2}^2 + x_i^2}, Z\right).
\end{align}

\begin{figure}
\centerline{{\includegraphics[width=5.5cm]{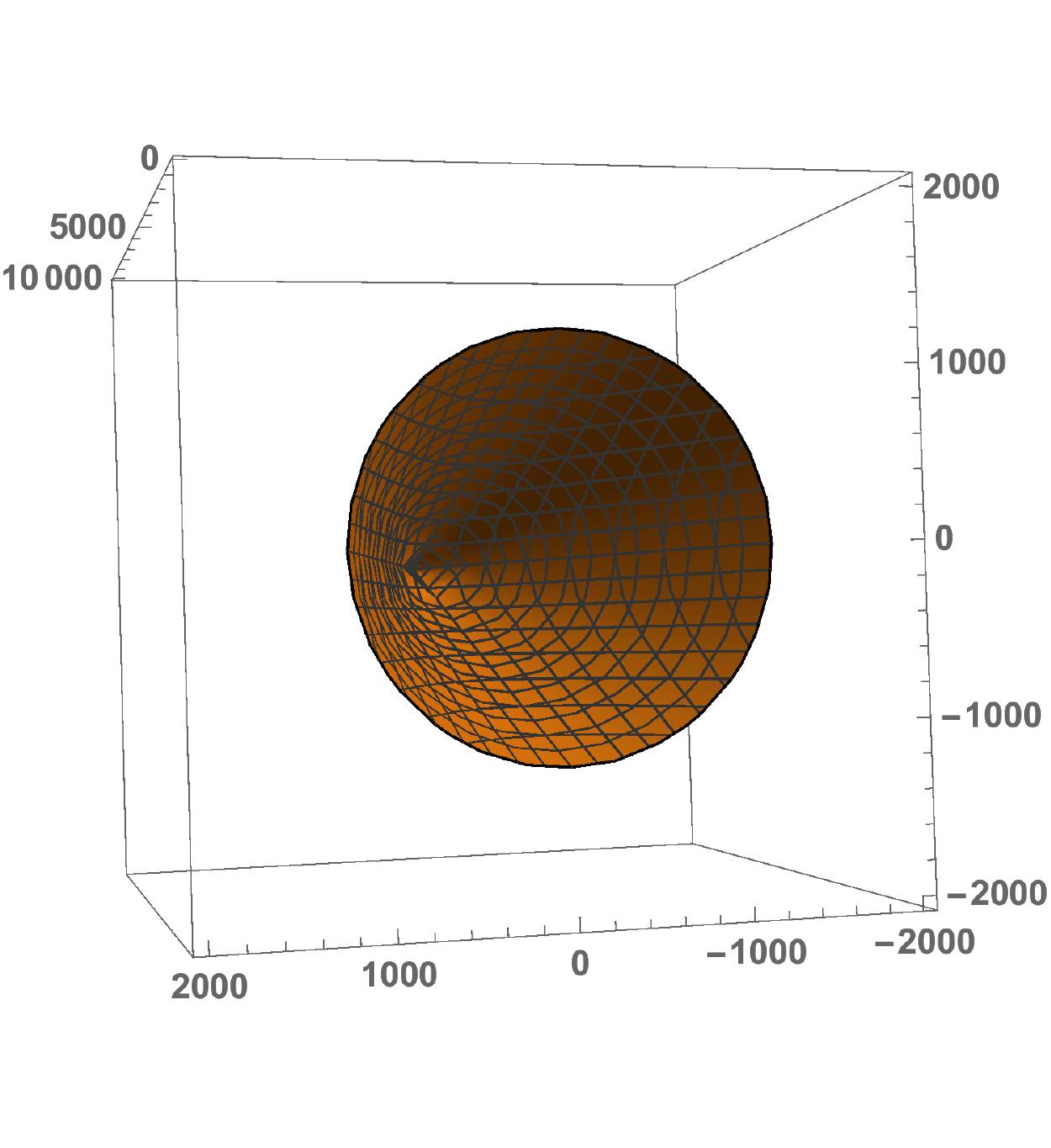}}\hfill{\includegraphics[width=5.5cm]{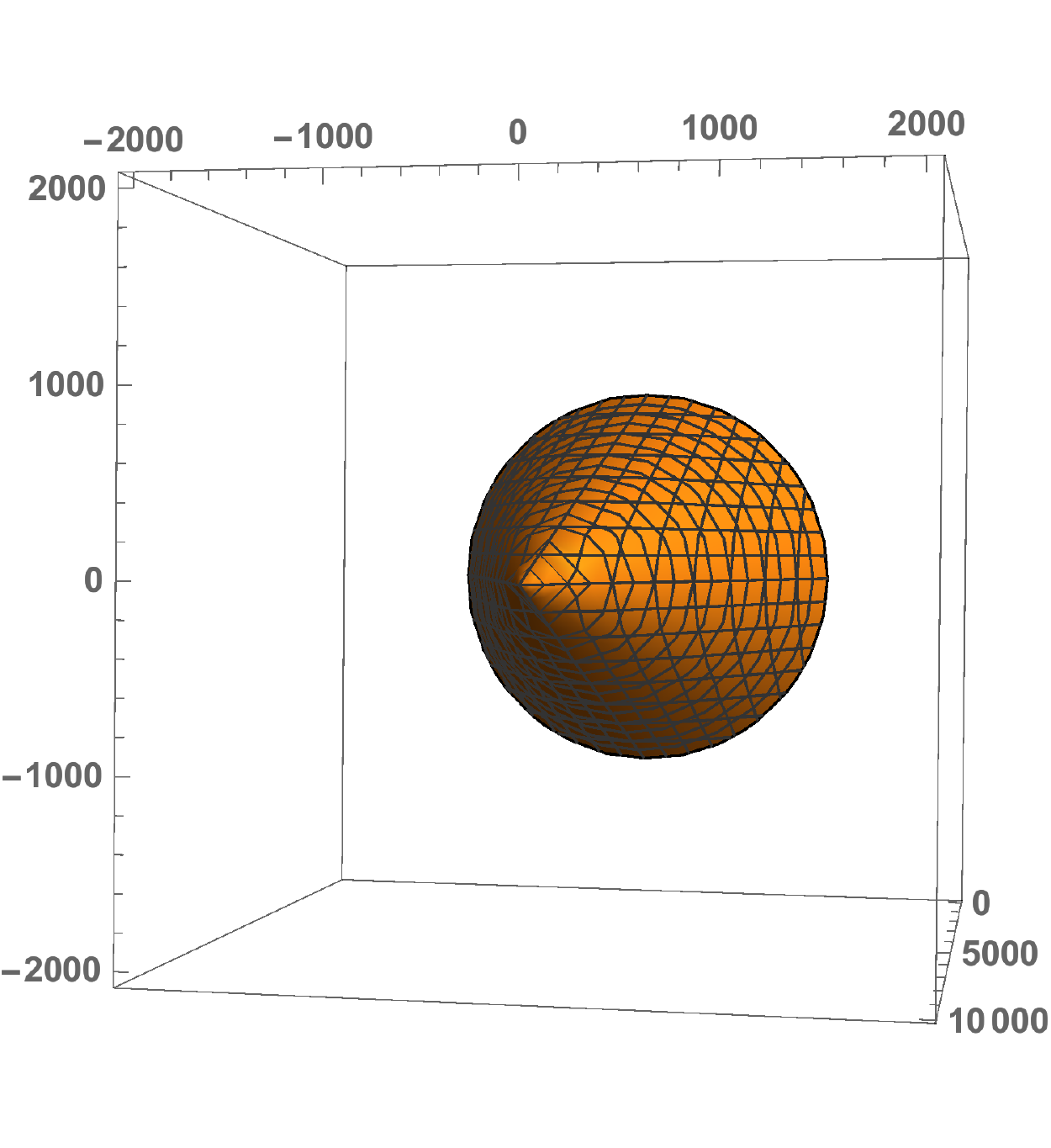}}\hfill{\includegraphics[width=5.5cm]{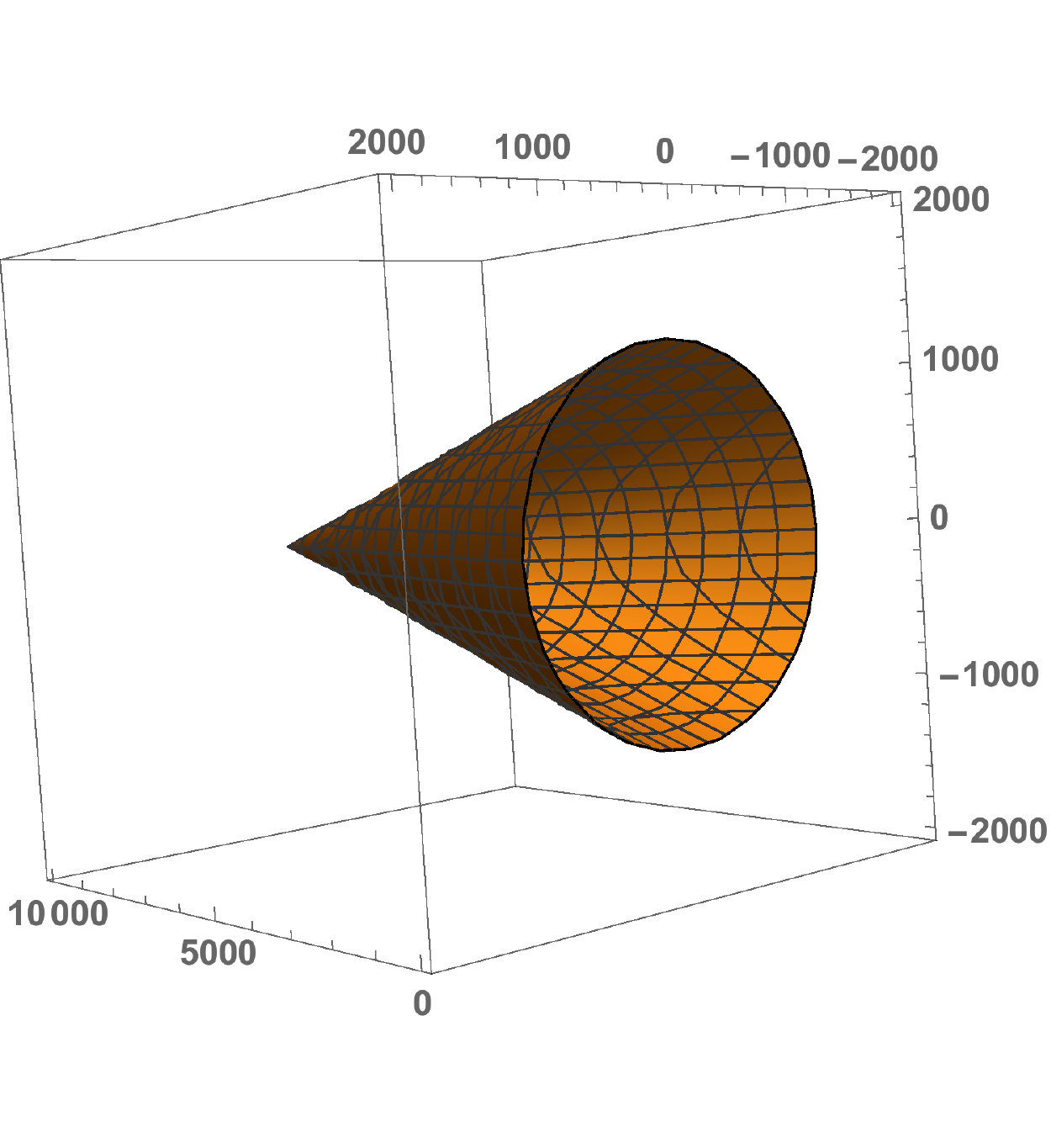}}}
\caption{Examples of the orientation of the jet with respect to the observer in the cases (i), (ii) and (iii), from left to right.}
\label{cones}
\end{figure}

\begin{figure}
\centerline{{\includegraphics[width=8cm]{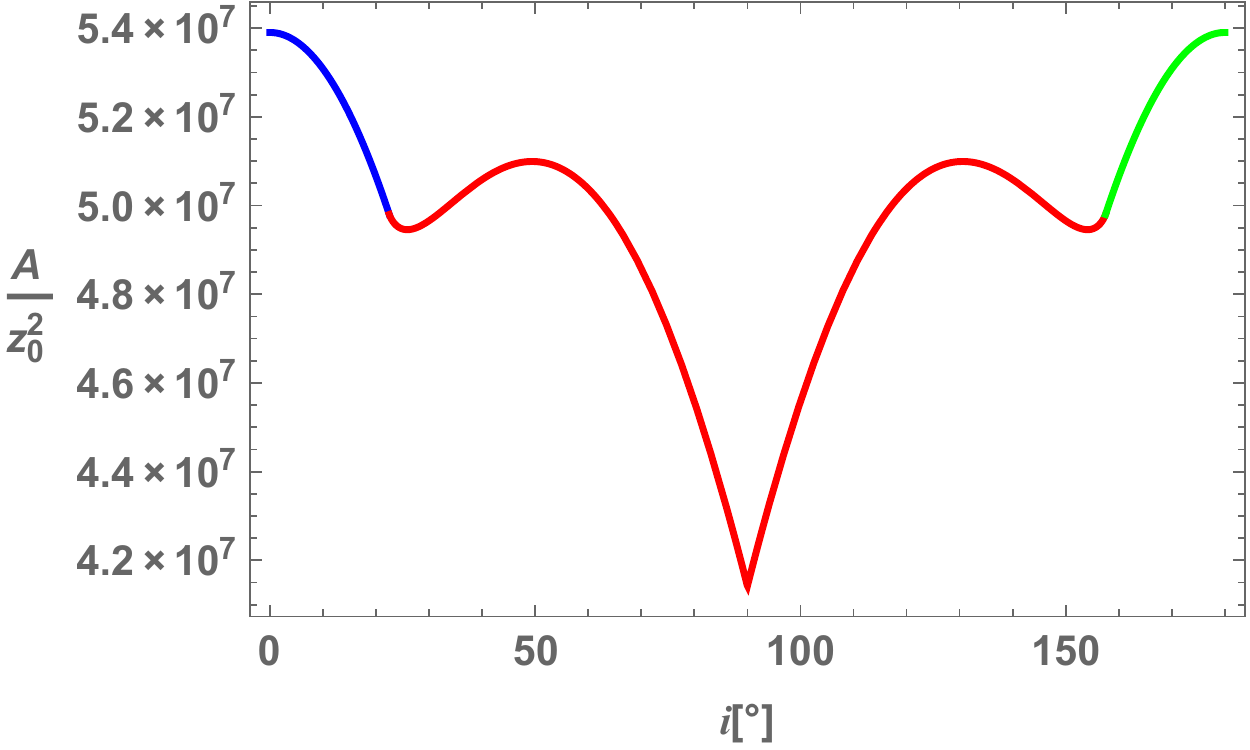}}{\hskip 0.5cm}{\includegraphics[width=8cm]{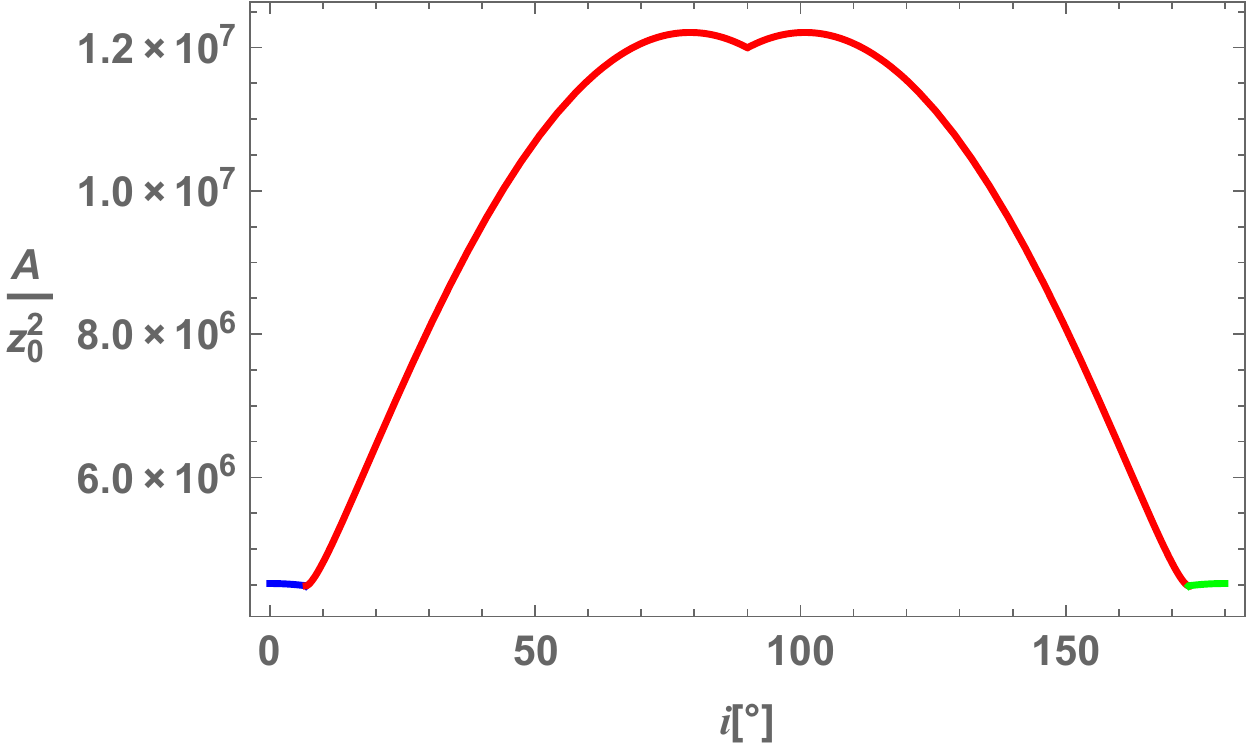}}}
\caption{Examples of the dependence of the projected area, $A$, on the viewing angle for $\theta_{\rm j}=22.5\degr$ (left) and $\theta_{\rm j}=6.84\degr$ (right; this value is the same as that used in Fig.\ \ref{angular}), both with $Z = 10^4$. The curves show the dependence for the case (i) at $i<\theta_{\rm j}$ (blue), case (iii) at $i \in (\theta_{\rm j},\upi-\theta_{\rm j})$ (red) and case (ii) at $i<\theta_{\rm j}$ (green).
}
\label{area}
\end{figure}

In the next step, we integrate the intensity over the jet projected area in the rotated coordinates. Depending on the orientation of the observer with respect to the jet, we have three different cases, illustrated in Fig.\ \ref{cones}. (i) $i<\theta_{\rm j}$, for which the projected area changes from a circle at $i=0$ but it contracts into an ellipse as $i$ gets larger. (ii) $\upi\geq i>\upi-\theta_{\rm j}$, which is very similar but the jet is viewed from the back. (iii) $i \in (\theta_{\rm j},\upi-\theta_{\rm j})$, in which case the projected area is shaped similarly to an ice cream cone. Two examples of the changes of the projected area with $i$ are shown in Fig.\ \ref{area}. In the cases (i) and (ii), the boundaries are easy to calculate; we can just apply the transformation of equation (\ref{rotated}) to the equation for a circle in the jet coordinates at height $Z$. This yields
\begin{align}
&\frac{{\rm d}F_{1}}{{\rm d}E} =\frac{1+z_{\rm r}}{m_{e}c^{2}D^{2}} \int_{-Z\tan \theta_{\rm j} \cos i}^{Z\tan \theta_{\rm j} \cos i}\int_{-\sqrt{Z^2 -(y_i/\cos i)^2}}^{\sqrt{Z^2 -(y_i/\cos i)^2}} I {\rm d}x_i {\rm d}y_i,\label{top_view}
\\
&\frac{{\rm d}F_{2}}{{\rm d}E} =\frac{1+z_{\rm r}}{m_{e}c^{2}D^{2}} \int_{-Z\tan \theta_{\rm j} \cos (i-\upi)}^{Z\tan \theta_{\rm j} \cos (i-\upi)}\int_{-\sqrt{Z^2 -[y_i/\cos (i-\upi)]^2}}^{\sqrt{Z^2 -[y_i/\cos (i-\upi)]^2}} I {\rm d}x_i {\rm d}y_i,\label{bottom_view}
\end{align}
in the cases (i) and (ii), respectively. We can see that these equations become identical to the flux given by equations (\ref{Fi0}) and (\ref{Fcj}) in the cases of $i = 0$ and $i = \upi$, respectively, for $Z\rightarrow \infty$.

\begin{figure}
\centerline{{\includegraphics[width=5.85cm]{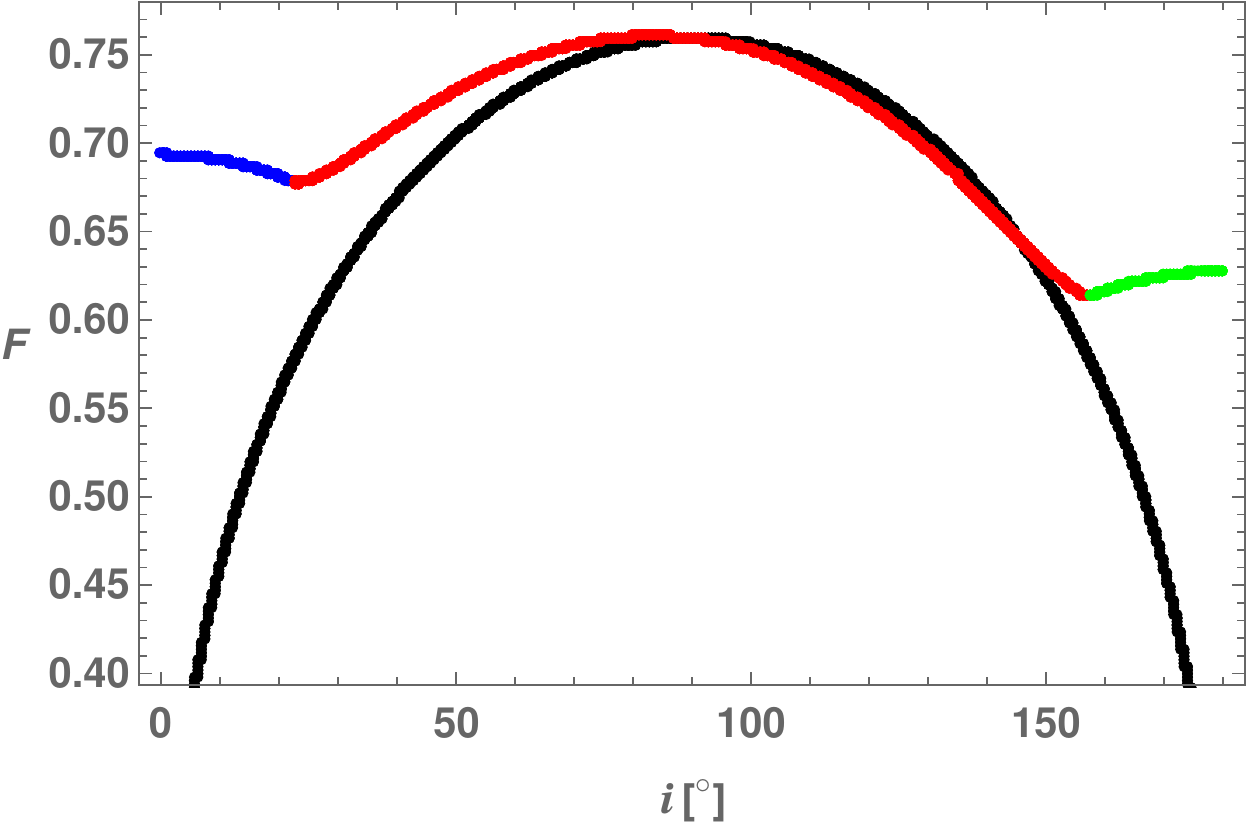}}{\hskip 0.01cm} {\includegraphics[width=5.85cm]{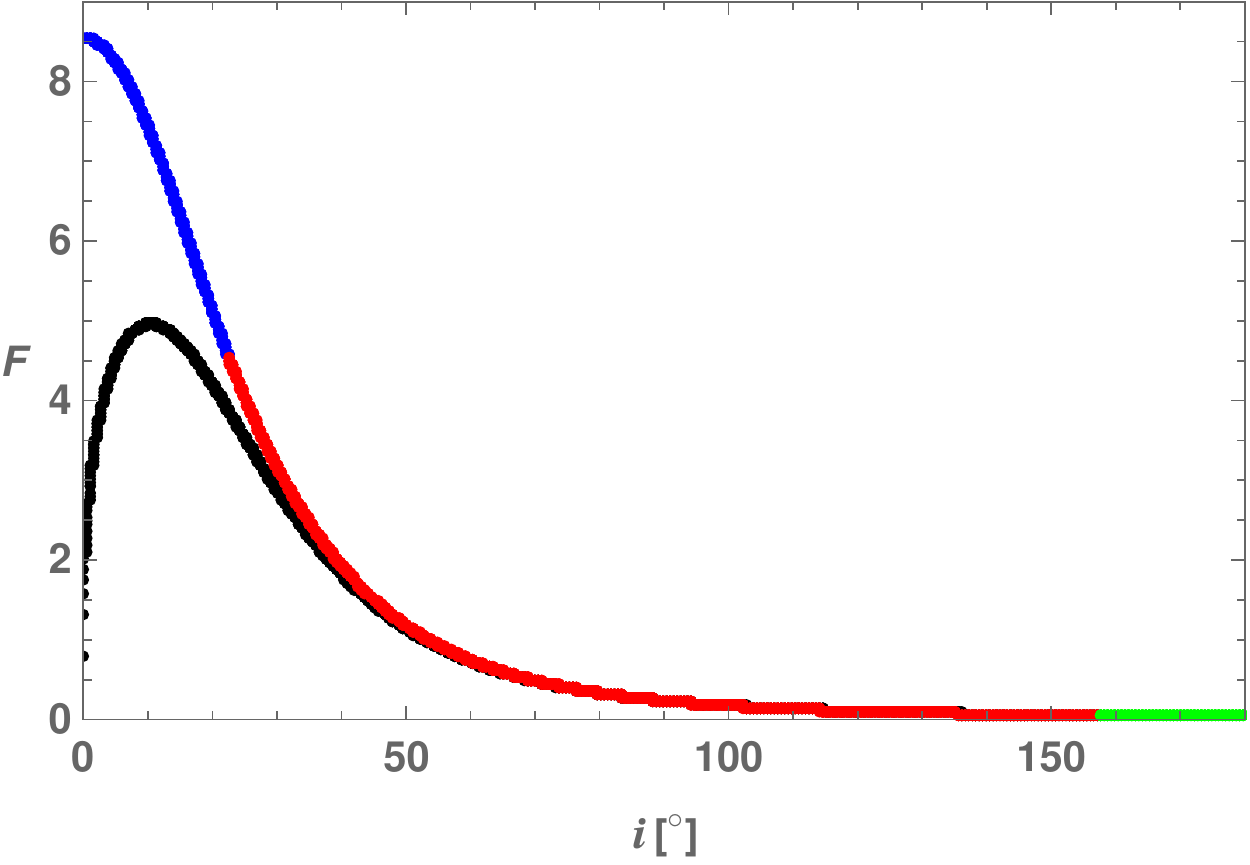}{\hskip 0.01cm}} {\includegraphics[width=5.85cm]{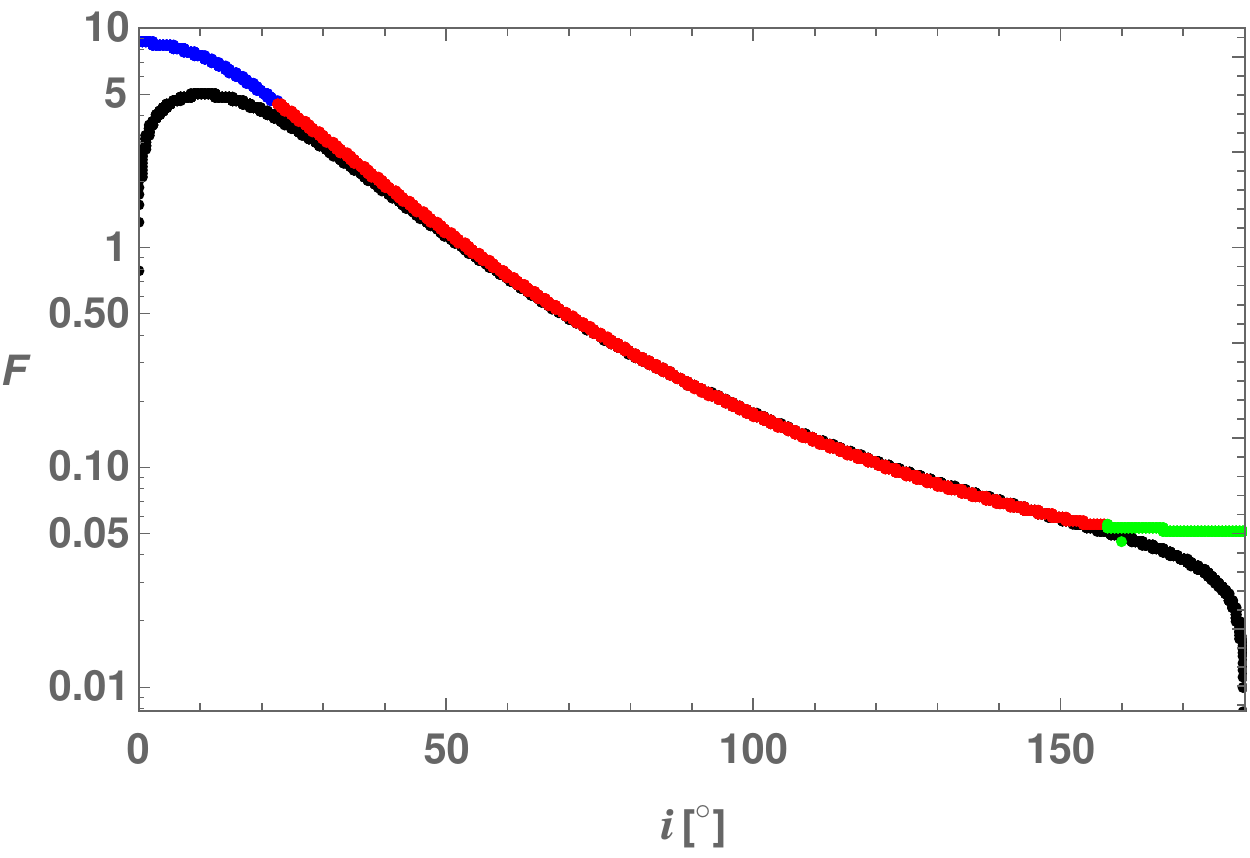}} }
\caption{The dependence of the flux on the viewing angle for $\theta_{\rm j} = 22.5\degr$, $S_{0} = \alpha_{0} = \epsilon = z_{0} = 1$, $p = 3$, $Z = 10^4$ (which is much larger than the value of $z$ corresponding to $\tau=1$, which implies the results being virtually independent of $Z$), and $\beta_{j} = 0$ (left) and 0.8 (middle and right). The middle and right panels have the linear and logarithmic vertical axes, respectively. The upper curves show the exact model for the case (i) at $i<\theta_{\rm j}$ (blue), case (iii) at $i \in (\theta_{\rm j},\upi-\theta_{\rm j})$ (red) and case (ii) at $i<\theta_{\rm j}$ (green). The lower (black) curves shows the cylindrical model for the flux, equations (\ref{int1}), (\ref{thick}). We see the two regimes connect smoothly in the exact model and the cylindrical approximation becomes more accurate for increasing viewing angle.}
\label{fluxes}
\end{figure}

In the case (iii), the integral must be split into two parts in order to accommodate the change of the domain from triangular to a part of an ellipse. Directly finding how the edge of the cone transforms is difficult but it is easy to find where the base of the jet is in the rotated coordinate system. We also know the shape of the ellipse and that the edge must intersect exactly once tangentially, so we can form the line using these two points. The intersections at the ellipse are given by $(v,u)$ and $(v,-u)$ in the $x_i$-$y_i$ plane, where
\begin{align}
&u = Z \cos i \cot i \tan^2 \theta_{\rm j},\\
&v = Z \tan \theta_{\rm j} \sqrt{1-\cot^2 i \tan^2 \theta_{\rm j}}.
\end{align}
The edge lines, giving some of the integration limits in equation (\ref{flux3}) below, are then given by,
\begin{align}
&p(y_i) = \frac{u y_i}{v \cos^2 i}-\frac{Z u \tan i}{v\cos^2 i},\\
&q(y_i) = \frac{-u y_i}{v\cos^2 i}+\frac{Z u \tan i}{v\cos^2 i}.
\end{align}
Past the intersection, the domain is the remaining part of the ellipse although now a modulus sign is needed to ensure that we are always integrating to the furthest point on the ellipse even when the viewing angle passes $\pi/2$ and $\cos i$ changes sign. This leads us to the final equation,
\begin{equation}
\frac{{\rm d}F_{3}}{{\rm d}E} =\frac{1+z_{\rm r}}{m_{e}c^{2}D^{2}} \int_{u}^{Z \sin i}\int_{p}^{q} I {\rm d}x_i {\rm d}y_i+\int_{-Z\tan (\theta_{\rm j}) |\cos i|}^{u}\int_{-\sqrt{Z^2 -(y_i/\cos i)^2}}^{\sqrt{Z^2 -(y_i/\cos i)^2}} I {\rm d}x_i {\rm d}y_i.
\label{flux3}
\end{equation}
This equation becomes identical to the flux of equation (\ref{int1}) in the case of $i = \upi/2$ for $Z\rightarrow \infty$. Fig.\ \ref{fluxes} shows two examples of this solution, and compares it to the cylindrical approximation, equations (\ref{int1}) and (\ref{thick}).

\label{lastpage}

\end{document}